\documentclass[twocolumn,twoside]{revtex4}
\usepackage{graphicx}
\usepackage{fancyhdr}
\usepackage{cancel,multirow,amsmath,pbox}
\begin{document}

\title{Signatures of generalized ALP interactions in SM decays of mesons}

%

\author{Subhajit Ghosh\footnote{Speaker}}
\affiliation{Department of Physics and Astronomy, University of Notre Dame, South Bend, IN 46556, USA}
\author{Triparno Bandyopadhyay}
\author{Tuhin S. Roy}
\affiliation{Department of Theoretical Physics, 
	Tata Institute of Fundamental Research,
	Mumbai 400005,  India.}

\begin{abstract}

 In addition to giving rise to spectacular new physics signals in the final states of meson decays, Axion-like-particles also induce modifications in the standard model decays of mesons. These `indirect' signatures can be parametrized as the modifications of the hadronic form factors and can be probed using meson decay width and decay distribution measurements. Starting with a generalized ALP Lagrangian, we demonstrate these effects for semileptonic Kaon decays and derived bounds using NA48/2 data. We also briefly discuss other indirect signatures such as modification of meson mass spectrum and `sum rules' comprised of meson decay amplitudes which show deviation in presence of ALP.
\end{abstract}

\maketitle

\thispagestyle{fancy}


\section{Introduction}
Proposed as a dynamical solution to the Strong CP problem~\cite{Peccei:1977hh}, Axion has emerged as a highly versatile tool for new physics model building (for review, see Ref.~\cite{Choi:2020rgn,Bauer:2017ris,Bauer:2021mvw}). Immense experimental efforts are ongoing to detect Axion or Axion-like-particle (ALP) through its interaction with quarks, leptons, and the gauge bosons. The majority of these experiments belong to the class of `direct' searches, where the ALP is either present in the initial state or is subsequently produced via the interaction of Standard Model particles. In the latter case, the final state ALP leaves its signature in the detector either decaying back into Standard Model (SM) particles or in the form of missing energy if its decay lifetime is beyond the experimental sensitivity.

Interestingly, there exists a completely independent way of searching for these particles by looking into the decays of mesons in SM final states. These `indirect' searches do not contain the ALP  in either the initial or the final state of the decay process. ALP, in general, has mass mixing and kinetic mixing interactions with the neutral mesons. Diagonalization of these mixings in the physical basis modifies the decay rates of mesons to the SM states. Thus deviations of these decay rates from the SM predictions can put independent constraints on the ALP couplings with the SM. Since the ALP does not take part in these decay processes, the bounds derived from indirect searches are independent of ALP lifetime or ALP decay channels.

In Ref.~\cite{Bandyopadhyay:2021wbb}, we demonstrated how precision measurements of semileptonic decays of charged Kaon can be used to put constraints on the ALP parameter space.  We further showed how these indirect searches complement the direct searches of ALP. We also construct `sum rules' using partial decay widths of mesons which can distinguish the effects of different classes of ALP operators.

\section{Generalized ALP Lagrangian}

In Ref.~\cite{Bandyopadhyay:2021wbb}, we start out with a `generalized' set of operators upto dimension $d =5$ in the electroweak basis involving the ALP and the quarks, where only the periodic symmetry of the ALP is taken into account. These operators are
 \begin{equation}
 \label{eq:1}
 \sum_{i=0}^8 \left( 
 C_{L}^i \mathcal{O}_{L}^i + C_{R}^i \mathcal{O}_{R}^i   
 + C_{LR}^i \mathcal{O}_{LR}^i  
 \right) 
 + C_{W} \mathcal{O}_{W}   
 + C_{Z} \mathcal{O}_{Z} 
 \; ,
 \end{equation}
 with $C_a$s being the \emph{real} Wilson coefficients for the operator of class $\mathcal{O}_a$. We tabulate the operators in Table.~\ref{tab:1}.
 \begin{table}
 	\centering
 	\begin{tabular}{|c|c|l|}
 		\hline
 		&&\\
 		\multirow{4}{*}{$\mathcal{O}_{L}^i$}
 		&\multirow{4}{*}{$\dfrac{1}{f_a} \partial_\mu a\:\overline{q}_L t^i \gamma^\mu q_L$}
 		&$i = 0,8:$ Allowed \\
 		&& $i = 1,2,4,5:$ break EM \\
 		&&$i = 1(1)7:$ break $SU(2)_W$\\
 		&& $i = 6,7:$ tree FCNC \\
 		&&\\
 		\hline
 		&&\\
 		\multirow{2}{*}{$\mathcal{O}_{R}^i$}
 		&\multirow{2}{*}{
 			$\dfrac{1}{f_a}\partial_\mu a\: \overline{q}_R t^i\gamma^\mu q_R$}
 		&
 		\\
 		&&\(i=0,3,8:\) Allowed\\
 		&&\\
 		\cline{1-2}
 		&&$i = 1, 2,4,5:$ break EM\\
 		\multirow{2}{*}{ $\mathcal{O}_{LR}^i $ }
 		&\multirow{2}{*}{
 			$\dfrac{a}{f_a}  \:  \overline{q}_L t^i  M q_R$
 		}
 		&$i = 6,7:$ tree FCNC\\
 		&&\\
 		&&\\
 		\hline
 		&&\\
 		$\mathcal{O}_{W} $
 		& $-\dfrac{a}{f_a}  \: \overline{q}_L 
 		Q^W \cancel{j}_\pm q_L$
 		
 		&\\
 		&& \\
 		\cline{1-2}
 		&& \\
 		$\mathcal{O}_{Z} $ 
 		&
 		\pbox{9 cm}{ $-\dfrac{a}{f_a}  \:\left( \overline{q}_L
 		Q_L^Z\; \cancel{j}_Z q_L \right. $\\$ \left.
 		+ \ \overline{q}_R Q_R^Z\; \cancel{j}_Z q_R \right)$}
 		&\\
 		&&\\
 		\hline
 	\end{tabular}
     \caption{The  dimension-5 ALP operators in the EW basis. The rightmost column indicates the generators allowed by symmetry and also the ones not considered by us. $t^i$s are the $ SU(3) $ generators and $ t^0 $ is the identity matrix. The $j^\mu_\pm$ and $ j^\mu_Z $ are current replacements of $ W_\pm^\mu $ and $ Z^\mu $, respectively.
 }
 \label{tab:1}
\end{table}
This framework includes operators $ \left(\mathcal{O}_W,\mathcal{O}_Z\right) $ where the ALP appears as a scalar and not a pseudo-scalar, allowing us to include models where the ALP may mix with CP even states.
In this study, we have only considered operators linear in $ a/f_a $ where $ f_a $ is the ALP decay constant.   

We match the quark Lagrangian in Eq.~\eqref{eq:1} with the meson Lagrangian using leading order $ SU(3) $ $ \chi $EFT. Operator  $\mathcal{O}_{LR}^i $ induces mass mixing between the neutral mesons and the ALP, whereas, $\mathcal{O}_{L}^i$ and $\mathcal{O}_{R}^i$ generate kinetic mixing. The Lagrangian, after basis diagonalization, captures all the essential physics of indirect signatures of the ALP.
\subsection{Modification of the form factors for $ K_{\ell 3} $ decays}
Due to the mixing of the ALP with the neutral mesons the form factors (FF) of the mesonic states get modified. In Ref.~\cite{Bandyopadhyay:2021wbb} we focussed on the \(K^+\to \pi^0\ell^+\nu\) (\(K^+_{\ell_3}\)) decays and studied the effects of ALP interaction on strangeness violating FF defined as:
\begin{subequations}
	\label{eq:FFDef}
	\begin{multline}
	\langle \pi^0(p_\pi)|\bar{s}\gamma_\mu u|K^{+}(p_{K})\rangle \\ 
	\equiv 
	\frac{1}{\sqrt{2}} \Big[ f_{+}^{K^{+}
		\pi^{0}}(q^2)\; Q_\mu 
	+ f_{-}^{K^{+}\pi^{0}}(q^2)\;q_\mu  
	\Big],
	\\
	\mathrm{where,}\quad Q^\mu= p_K^\mu+p_\pi^\mu;\quad q_\mu=
	p_K^\mu-p_\pi^\mu.
	\end{multline}
\end{subequations} 
At leading order $ \left(\mathcal{O}(p^2)\right) $ $ \chi $EFT, the amplitude of $ K_{\ell _3} $ process in terms of the FF is written as;
\begin{multline}
\label{eq:ampDef}
\mathcal{A} =  G_F  V_{\bar{s}u} 
\left[ \tilde{f}_{+}^{K^{+} \pi^{0}}(0)\, Q_\mu +  
\tilde{f}_{-}^{K^{+} \pi^{0}}(0)\, q_\mu  \right]  \\ \times
\bar{u}_\nu \gamma^\mu \frac{1}{2}\left(1 - \gamma_5 \right) v_\ell.
\end{multline}
Here $ G_F $ is the Fermi constant, $ V_{\bar{s}u} $ is the CKM matrix elements and $ \tilde{f}_\pm^{K^{+} \pi^{0}}(0) $ are the `effective' FF which include additional contributions for the hadronic amplitude beyond the modifications of $ f_{\pm}^{K^{+} \pi^{0}} $~\cite{Bandyopadhyay:2021wbb}.
The ALP-induced modifications of the FF are parametrized as;
\begin{align}
\label{eq:albetDef}
\begin{split}
&\tilde{f}_{+}^{K^{+} \pi^{0}}(0) = 
\alpha^{(0)}_{K^+\pi_0}  + \xi^2
\left(\alpha^{(2)}_{K^+\pi_0} 
+ i \tilde{\alpha}^{(2)}_{K^+\pi_0} \right)\;, \\
&\tilde{f}_{-}^{K^{+} \pi^{0}} (0) = 
\beta^{(0)}_{K^+\pi_0} +  \xi^2 
\left(\beta^{(2)}_{K^+\pi_0} 
+ i \tilde{\beta}^{(2)}_{K^+\pi_0}\right), \text{with}~~\\
&
\alpha^{(0)}_{K^+\pi_0}= 1 - \sqrt{3}\, \epsilon, \\
& \alpha^{(2)}_{K^+\pi_0}= 
-\frac{C_3}{8}( C_{LR}^3 - C_R^3 +2\sqrt{3} (C_{LR}^8 - C_R^8)),\\
&\tilde{\alpha}^{(2)}_{K^+\pi_0}= 
-\frac{1}{2} C_3 C_W, \ \text{and}~~ \\&
\beta^{(0)}_{K^+\pi_0}= 0,\;
\beta^{(2)}_{K^+\pi_0}=-\frac{\sqrt{3}}{4}C_3 C_L^8,\;
\tilde{\beta}^{(2)}_{K^+\pi_0}=\frac{1}{2}C_3 C_W,\\
&\text{where}~\epsilon = {\sqrt{3} \over 4}{m_u - m_d \over m_s - (m_u+m_d)/2},\\
&\text{and}~C_{3} = C_{LR}^{3}-C_R^{3}.
\end{split}						
\end{align}
\begin{figure}
	\centering
	\includegraphics[width= \linewidth]{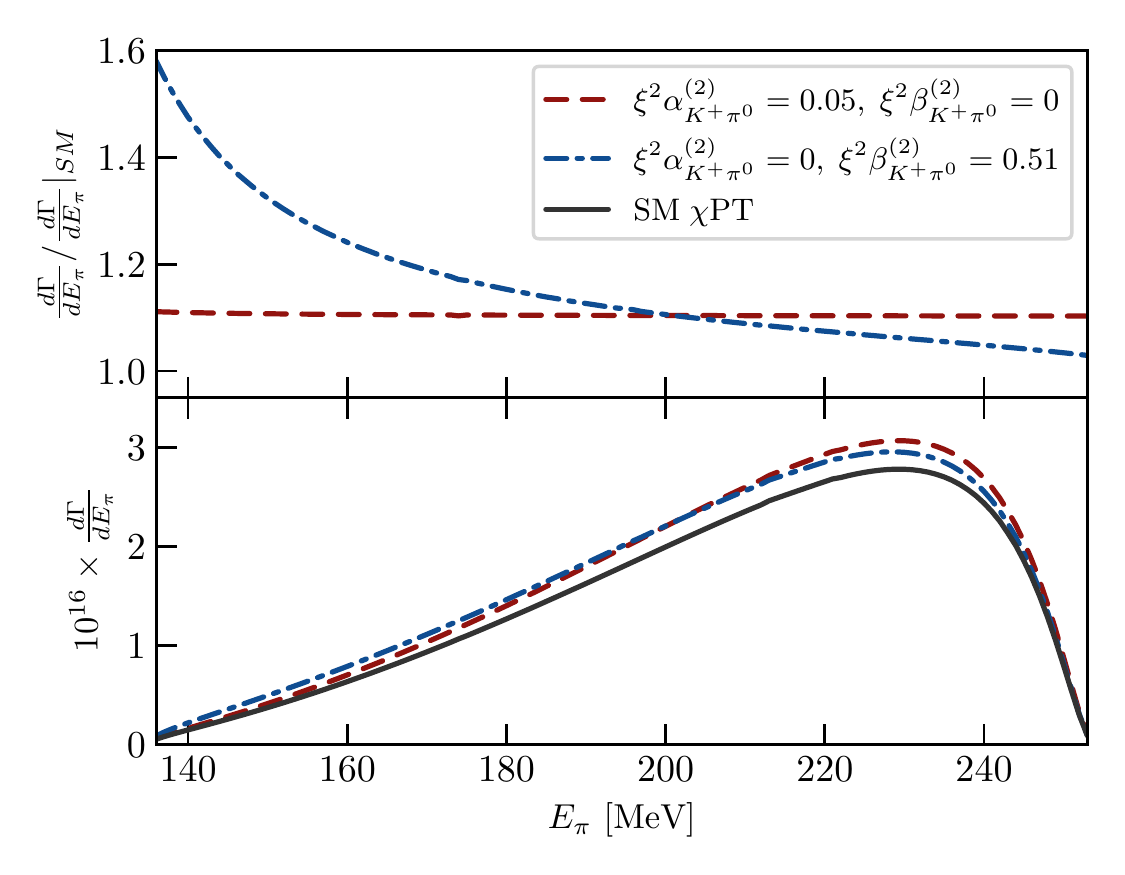}
	\caption{
		Bottom: Differential spectrum of \(K^+_{\mu_3}\), w.r.t.
		\(E_{\pi}\) for \(\xi^2\alpha^{(2)}_{K^+\pi^0}= 0.05\) (dashed
		red) and \(\xi^2 \beta^{(2)}_{K^+\pi^0} = 0.51\) (dash-dotted
		blue).  The values of the parameters are chosen such
		that the total decay width is same for both graphs. The SM spectrum is shown in black.
	}
	\label{fig:alpha-beta-comp-wSM}
\end{figure}
Here $ \alpha^{(0)} $  and $ \beta^{(0)} $ correspond to the SM contributions at leading order in $ \chi $EFT. 
The imaginary pieces of the ALP-induced modifications of the FF contribute at $ \mathcal{O}(\xi^4) $, whereas, the real parts, which interfere with the SM contribution, produce effects at $ \mathcal{O}(\xi^2) $. Therefore, we drop the imaginary contributions from our calculation.
We incorporate the higher order $ \chi $EFT effects in this formalism, which introduce momentum dependence in the FF, via   
\begin{align}
\mathrm{Re}\left( \tilde{f}_{+}^{K^{+} \pi^{0}}(t)\right) & \simeq
\left[
1 + \xi^2  
\frac{\alpha^{(2)}_{K^+\pi^0}}{\alpha^{(0)}_{K^+\pi^0}}
\!\right] f_{+,\,
	\mathrm{SM}}^{K^{+} \pi^{0}}(t)\;, \\
\mathrm{Re}\left( \tilde{f}_{-}^{K^{+} \pi^{0}}(t)\right) &\simeq 
\left[
1 + \xi^2  
\frac{\beta^{(2)}_{K^+\pi^0}}{\delta\beta^{(0)}_{K^+\pi^0}}
\!\right] f_{-,\,
	\mathrm{SM}}^{K^{+} \pi^{0}}(t)\;.
\end{align}
Here $ \delta\beta^{(0)}_{K^+\pi^0} $ is the higher order $ \chi $EFT contribution to $ \beta^{(0)}_{K^+\pi^0} $.
In writing the above equations, we have ignored the ALP-induced corrections to the `slope parameters', which are subleading in nature.
\begin{figure*}
	\centering
	\includegraphics{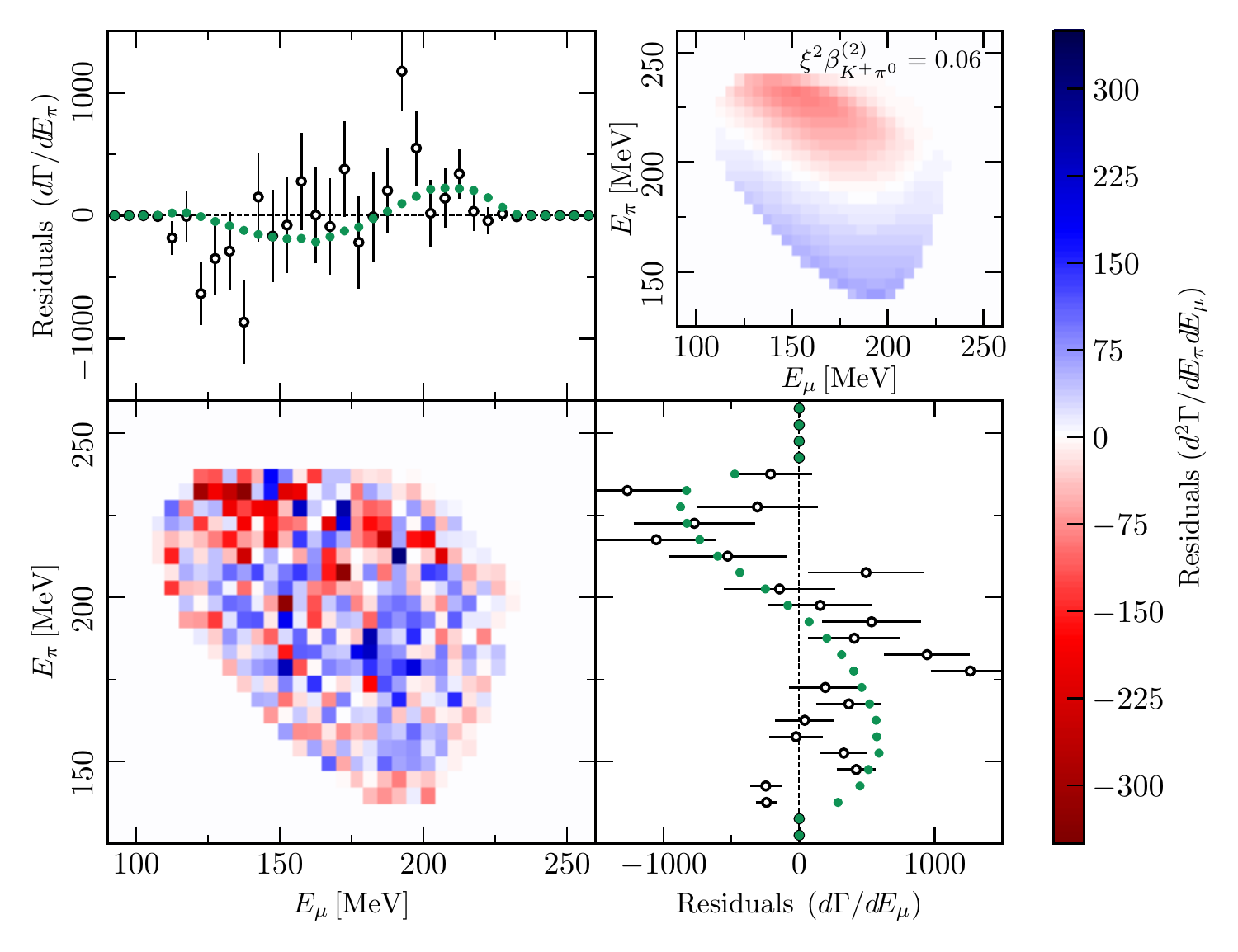}
	\caption{Bottom Left: Binned Dalitz plot for the residual (the difference between the data and the SM expectation). Top Right: Expected Dalitz plot signal for \(\xi^2
		\beta^{(2)}_{K^+\pi^0} = 0.06\). The bottom-right and top-left are the marginalized residuals (with errorbar) and signals with
		respect to \(E_\mu\) and \(E_\pi\), respectively.}
	\label{fig:exp}
\end{figure*}
Fig.~\ref{fig:alpha-beta-comp-wSM} shows the modification of the $ K_{\ell _3} $ differential rate. $ \xi ^2\alpha^{(2)} $, which modifies the dominant $\tilde{f}_{+}^{K^{+} \pi^{0}}$ FF, results in almost constant modification with respect to SM. Whereas $ \xi^2\beta^{(2)} $ induces signification distortion of the differential distribution. Note that, for the same values of the ALP parameters, the modification due to $ \xi^2\beta^{(2)}  $ is much smaller, since its effect is suppressed by the final state lepton mass. 
\section{Constraints}

In this section, we depict the current constraints of the ALP parameters from $ K_{\ell _3} $ decays derived using the following datasets:
\begin{itemize}
	\item The decay distribution measurements for $ K_{\mu _3} $ and $ K_{e _3} $ by the NA48/2 collaboration~\cite{Lazzeroni:2018glh},
	\item Measured total decay widths of  the above two decays~\cite{Zyla:2020zbs}.
\end{itemize}
We compare the theoretical SM expectations of the decay rates and the decay distributions with the above experimental datasets.
For the theoretical SM calculation, we use the FF parameters obtained from lattice computations by the European twisted mass collaboration~\cite{Carrasco:2016kpy}. We include the uncertainties of and the correlations among the theoretical values of the FF parameters in our computation~\cite{Bandyopadhyay:2021wbb,Carrasco:2016kpy}.

In Fig.~\ref{fig:exp} we compare the theoretical expectation with the measured data for $ K_{\mu 3} $ decay. The figure shows the residual Dalitz plot distribution which is the difference between the data and the theoretical calculation. We also show the 1D residual plots projected along $ E_\mu $ and $E_\pi$ containing the experimental errorbar. The 1D residual with respect to $ E_\mu $ shows a slight systematic excess which can be fitted using \(\xi^2
\beta^{(2)}_{K^+\pi^0} = 0.06\). However, the excess is not significant and becomes consistent with the SM when the theoretical uncertainties of the FF are taken into account.

\begin{figure}
	\centering
	\includegraphics[width=\linewidth]{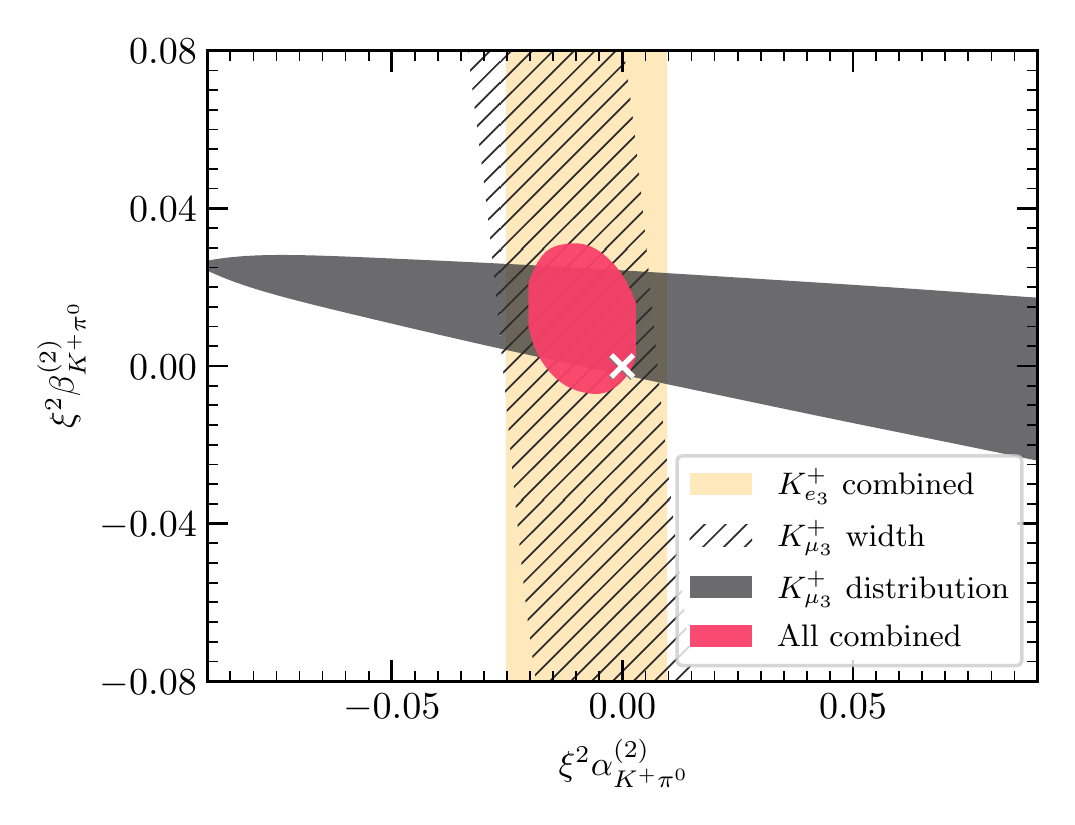}
	\caption{The allowed regions for the ALP parameters in
		the \(\xi^2 \alpha^{(2)}_{K^+\pi^0}\) -
		\(\xi^2\beta^{(2)}_{K^+\pi^0}\) plane at 95\% C.L. The black contour shows the constrains from the $ K_{\mu 3} $ distribution measurement. The yellow and the hached regin are the constrains derived from the total decay rate of $ K_{e 3} $ and $ K_{\mu 3} $, respectively. The red region shows the combined constrain. The white cross denotes the SM point.
	}
	\label{fig:alpha-beta-contour}
\end{figure}
\begin{table}
	\centering
	\renewcommand{\arraystretch}{1.4}
	\begin{tabular}{| c| c | c| }
		\hline
		&  \pbox{12 cm}{\(\xi^2\beta^{(2)}_{K^+\pi^0} \) \\ \(  \left(\xi^2 \alpha^{(2)}_{K^+\pi^0} =
		0\right)\)} & \pbox{12 cm}{\(\xi^2\alpha^{(2)}_{K^+\pi^0}\) \\ \(
		\left(\xi^2\beta^{(2)}_{K^+\pi^0} = 0\right)\)}\\
		\hline
		\hline
		\(K_{\mu_3}^+\) & $ [-0.006, 0.026]$  &    $ [-0.021, 0.007]$\\
		\hline
		\(K_{\mu_3}^++K_{e_3}^+\) & $ [-0.006,0.026] $& $    [-0.018,0.003]$\\
		\hline
	\end{tabular}
	\caption{The 95\% confidence limits on 
		\(\xi^2\alpha^{(2)}_{K^+\pi^0}\) and 
		\(\xi^2\beta^{(2)}_{K^+\pi^0}\) from the \(K_{\mu_3}^+\) decay analysis and the combined $K_{\mu_3}^++K_{e_3}^+$
		analysis.
	}
	\label{tab:beta-alpha-fixed}
	\renewcommand{\arraystretch}{1}
\end{table}
We performed a $ \chi^2 $ analysis using all the datasets quoted at the beginning of the section. The contours at 95\% C.L. in the \(\xi^2 \alpha^{(2)}_{K^+\pi^0}\) -
\(\xi^2\beta^{(2)}_{K^+\pi^0}\) plane are shown in Fig.~\ref{fig:alpha-beta-contour}. The limits at 95\% C.L. from the analysis are shown in Table.~\ref{tab:beta-alpha-fixed}. The constraints on \(\xi^2\beta^{(2)}_{K^+\pi^0}\) from the total width measurements of both $ K_{\mu_3} $ and $ K_{e_3} $ are substantially weaker owing to the associated lepton mass suppression. However, from the decay distribution measurements,  the constraint on  \(\xi^2\beta^{(2)}_{K^+\pi^0}\) is much stronger compared to \(\xi^2\alpha^{(2)}_{K^+\pi^0}\). This is due to the fact that the decay distribution measurement of NA48/2 is mostly sensitive to the shape modification of the spectrum, not the normalization ( i.e, not sensitive to the overall scaling of total decay rate). Since, \(\xi^2\beta^{(2)}_{K^+\pi^0}\) induces a dissimilar momentum dependence with respect to the SM (as shown in Fig.~\ref{fig:alpha-beta-contour}), the constraint is stricter in this case~\cite{Bandyopadhyay:2021wbb}. Note that, $ K_{e_3} $ distribution measurement does not provide any meaningful constraint on the ALP parameter space due to the electron mass suppression. The combined constraint from both decay width and distribution measurements is shown in red and is consistent with the SM at 95\% C.L.
\section{Other indirect signatures}

In addition to the FF modifications, ALP also induces a number of interesting indirect signatures. Due to the mass and kinetic mixing with the ALP, the mass spectra of the mesons deviate from the SM values. These deviations can be constrained using theoretically clean mass ratio observables, such as the well-known Gell-Mann-Okubo mass relation and the charged and neutral pion mass difference ratio. In Ref.~\cite{Bandyopadhyay:2021wbb} we sketch the resulting constraints from these observables.

The constraints derived using the indirect method largely complement the ones derived via direct observations. Using the generalized Lagrangian in Eq.~\ref{eq:1}, it can be shown that for certain values of the ALP parameters the direct decay amplitude for \(K^+\to a\ell^+\nu\)  goes to zero. Although the indirect signature still persists in those limits, i.e, the amplitude for  \(K^+\to \pi^0\ell^+\nu\) shows deviation from the SM expectation~\cite{Bandyopadhyay:2021wbb}. Thus, indirect methods can shine light on the direct detection blind spots generally known as `pion-phobia'.

We can also construct `sum-rules' using SM decay widths of mesons which provide a unique way of identifying the presence and nature of ALP interactions in meson Lagrangian. As shown in Ref.~\cite{Bandyopadhyay:2021wbb} the deviations from these sums from unity point towards the presence of ALP coupling with mesons, whereas, the signs of the deviations can be associated with different classes of ALP operators. Therefore, these sums, which are constructed out of low-energy observables, can shed light on the UV nature of the ALP physics.

\section{Conclusion}

Indirect searches open up a hitherto unexplored avenue of ALP phenomenology. The presence of a large number of indirect channels and the complementarity with the direct searches make this a viable approach to looking for ALP signatures. With the improvement in the theoretical predictions of the FF and high precision data from low-energy experiments, these bounds will improve significantly in near future. Similar analysis involving other indirect channels, such as heavy meson decays and tau decays, will be able to put further constrain on the ALP parameter space.

\begin{acknowledgments}
	
	The authors thank Dmitry Madigozhin for providing them with the source of the NA48/2 dataset.
	The research of SG is supported by the NSF grant PHY-2014165.

\end{acknowledgments}

\bigskip 
\bibliography{Axions}

\end{document}